\documentclass[onecolumn,showpacs,preprintnumbers,amsmath,amssymb,APSl,nofootinbib,prd]{revtex4}
\usepackage{bm}
\usepackage{hyperref}
\usepackage{mathrsfs}
\usepackage{color,graphicx,graphics}
\usepackage[all]{xy}
\usepackage{epsfig,subfigure}
\usepackage{latexsym,amssymb,amsmath,amsfonts} 
\usepackage[english, spanish, portuguese]{babel} 
\usepackage[OT1]{fontenc}
\usepackage[latin1]{inputenc}
\usepackage{makeidx}

\definecolor{red}{rgb}{1,0,0}

\def\p{\partial}

\def\<{\leftarrow}
\def\>{\rightarrow}

\def\O{\Omega}
 
\def\a{\alpha} \def\b{\beta}  \def\d{\delta} \def\e{\epsilon}
\def\m{\mu} \def\n{\nu} \def\r{\rho} \def\s{\sigma} \def\l{\lambda} 
\def\k{\kappa}\def\G{\Gamma}

\def\q{\quad}

\def\({\left(}  \def\){\right)}

\newcommand{\bi}{\begin{itemize}} 		\newcommand{\ei}{\end{itemize}}
\newcommand{\benu}{\begin{enumerate}} \newcommand{\enu}{\end{enumerate}}
\newcommand{\bd}{\begin{dinglist}{0}}     \newcommand{\ed}{\end{dinglist}}
\newcommand{\bfig}{\begin{figure}[htbp]}  \newcommand{\efig}{\end{figure}}
       			
\newcommand{\bc}{\begin{center}} 	\newcommand{\ec}{\end{center}}
\newcommand{\be}{\begin{equation}} 	\newcommand{\ee}{\end{equation}}
\newcommand{\bsub}{\begin{subequations}}  \newcommand{\esub}{\end{subequations}}
\newcommand{\ben}{\begin{eqnarray}} 	\newcommand{\een}{\end{eqnarray}}
\newcommand{\ba}[1]{\begin{array}{#1}} 	\newcommand{\ea}{\end{array}}
\newcommand{\bea}{\begin{equation}\begin{array}{rcl}} \newcommand{\eea}{\end{array}\end{equation}}

\newcommand{\Zt}{\tilde{Z}}
\newcommand{\Omegah}{\hat{\Omega}} 
\newcommand{\mR}{{\mathcal R}}

\begin{document}
\title{The role of torsion in projective invariant theories of gravity with non-minimally coupled matter fields}

\author{V. I. Afonso$^{a,d}$} \email{viafonso@df.ufcg.edu.br}
\author{Cecilia Bejarano$^{b,d}$} \email{cbejarano@iafe.uba.ar}
\author{Jose Beltr\'{a}n Jim\'{e}nez$^c$} \email{jose.beltran@cpt.univ-mrs.fr}
\author{Gonzalo J. Olmo$^{d,e}$} \email{gonzalo.olmo@uv.es}
\author{Emanuele Orazi$^{f,g}$} \email{orazi.emanuele@gmail.com}
\affiliation{$^{a}$Unidade Acad\^emica de F\'\i sica, Universidade Federal de Campina
Grande, 58109-970 Campina Grande, PB, Brazil}
\affiliation{$^{b}$Instituto de Astronom\'ia y F\'isica del Espacio (IAFE, CONICET-UBA),
Casilla de Correo 67, Sucursal 28, 1428 Buenos Aires, Argentina.}
\affiliation{$^{c}$Aix Marseille Univ, Universit\'e de Toulon, CNRS, CPT, Marseille, France.}
\affiliation{$^{d}$Departamento de F\'{i}sica Te\'{o}rica and IFIC, Centro Mixto Universidad de Valencia - CSIC. Universidad de Valencia, Burjassot-46100, Valencia, Spain.}
\affiliation{$^{e}$Departamento de F\'isica, Universidade Federal da
Para\'\i ba, 58051-900 Jo\~ao Pessoa, Para\'\i ba, Brazil.}
\affiliation{$^{f}$International Institute of Physics,
Federal University of Rio Grande do Norte,
Campus Universit\'ario-Lagoa Nova, Natal-RN 59078-970, Brazil}
\affiliation{$^{g}$Escola de Ci\^encia e Tecnologia,
Universidade Federal do Rio Grande do Norte,
Caixa Postal 1524, Natal-RN 59078-970, Brazil}

\date{\today}
\begin{abstract}
We study a large family of metric-affine theories with a projective symmetry, including non-minimally coupled matter fields which respect this invariance. The symmetry is straightforwardly realised by imposing that the connection only enters through the symmetric part of the Ricci tensor, even in the matter sector. We leave the connection completely free (including torsion) and obtain its general solution as the Levi-Civita connection of an auxiliary metric, showing that the torsion only appears as a projective mode. This result justifies the widely used condition of setting vanishing torsion in these theories as a simple gauge choice.
We apply our results to some particular cases considered in the literature like the so-called Eddington-inspired-Born-Infeld theories among others. We finally discuss the possibility of imposing a gauge fixing where the connection is metric compatible and comment on the genuine character of the non-metricity in theories where the two metrics are not conformally related.
\end{abstract}
\maketitle

\section{Introduction}

The remarkable properties of Born-Infeld electromagnetism \cite{BIE}, originally aimed at resolving divergences of point-like charged particles, motivated the search of a similar route to resolve the singularities of General Relativity \cite{D&G1998}. Among the different proposals, the so-called Eddington-inspired-Born-Infeld (EiBI) theory \cite{BIg} has attracted a lot of attention in the last years due to its extraordinary ability to get rid of cosmological and black hole singularities and numerous works have been devoted to constrain the model using different types of observations \cite{BIg-app}. Extensions and modifications of that model also lead to interesting results in cosmological and black hole scenarios  \cite{oor14,Makarenko:2014lxa,Jimenez:2015caa,Bambi:2016xme}. The exploration of these theories showed that their natural habitat is the framework of metric-affine geometries and precisely this formulation permitted the mentioned progress (see \cite{BeltranJimenez:2017doy} for a review). The reason for the necessity of considering these theories in the metric-affine approach is the avoidance of ghost-like instabilities that otherwise would be present in the metric formulation of theories with non-linear curvature terms in the action \cite{Stelle:1977ry}. The metric-affine (sometimes also called Palatini) formulation is characterised by unlocking the affine structure and disentangle it from the metric structure, which amounts to assuming that the geometry is not Riemannian {\it a priori}, but of metric-affine type, where the metric and the connection are regarded as fully independent objects. The spirit of this approach is that only the resulting field equations should specify the full geometrical structure of the spacetime and, in particular, the relation between the metric and the connection with the matter fields. In this regard, it must be noted that the EiBI theory has been systematically analysed for a constrained family of connections by assuming a vanishing torsion tensor\footnote{We refer here to the most widely explored Born-Infeld inspired theories of gravity, being the EiBI model the paradigmatic example. A noteworthy exception is the class of Born-Infeld theories based on the teleparallel formulation of gravity, where the torsion is actually the fundamental object \cite{Ferraro:2008ey,Fiorini:2009ux,Ferraro:2009zk,Fiorini:2013kba,Fiorini:2015hob,Fiorini:2016zrt}.}. Given the growing interest in this theory and its applications, we find it convenient a careful analysis of the role of torsion in its field equations and, with a little extra effort, extend the analysis to a much larger class of affine theories. This issue has been treated with care in \cite{BeltranJimenez:2017doy} for minimally coupled fields, which corresponds to the case in many practical applications. In the present work we will extend the analysis to more general cases that include non-minimally coupled matter fields. We will see that the structure of the equations remains essentially the same, but some differences arise that could have interesting phenomenological consequences.

The role of torsion in metric-affine theories of gravity has been previously considered in the literature in the context of e.g. $f(\mR)$ and $f(\mR, \mR_{\mu\nu}\mR^{\mu\nu})$ theories of gravity  \cite{Capozziello:2007tj,Capozziello:2008yx,Capozziello:2008kb,Capozziello:2009mq,Sotiriou:2009xt,Capozziello:2012gw,Olmo:2013lta}. The case of $f(\mR)$ theories is particularly relevant to our discussion because a degeneracy between different formulations of those theories was observed when torsion was explicitly considered in the dynamics \cite{Capozziello:2007tj,Sotiriou:2009xt,Olmo:2011uz}. To make a long story short, one can say that $i)$ the torsionless Palatini formulation of $f(\mR)$ theories and $ii)$ the metric compatible ($\nabla^\Gamma_\mu g_{\alpha\beta}=0$) formulation of these theories with torsion yield the same field equations. This result, non-trivial at first sight, suggests an intimate relation between torsion and non-metricity in metric-affine theories of gravity. The question whether an analogous relation also exists in the EiBI theory is an issue that we will explicitly address and clarify in this work. As we will see, the non-metricity structure of the EiBI theory is much richer than in the $f(\mR)$ case while their torsional properties are much closer. As a consequence, there is no possible trade between the torsional and the non-metricity properties, making it impossible to find a metric-compatible representation of the theory with torsion. This result roots in the existence of a projective symmetry in the theory and the fact that the torsion only enters as a projective mode which, then, can be gauged away. This symmetry will also explain the different non-metric structure of $f(\mR)$ and EiBI as a consequence of the existence of a gauge where the non-metricity vanishes. 

An important step forward that we make in the present work is the inclusion of non-minimally coupled matter fields. However, we will handle this with care. Since the projective invariance proves to be a crucial ingredient, we will only allow non-minimal couplings respecting this symmetry. The most straightforward procedure to realise it is by simply allowing non-minimal couplings only to the symmetric part of the Ricci tensor (which is exactly projective invariant). Although this might seem like a severe restriction, it is actually quite natural and permits a large class of couplings. In fact, very much like the metric-affine formalism opens up the possibility for much more general gravitational actions than the metric formalism allows, the same applies for non-minimal couplings. Within the metric formalism, it is known that derivative couplings of scalar fields to the curvature must be of a very specific form in order to avoid ghosts \cite{nonminimalscalar}. For vector fields, already non-derivative couplings to the curvature must be carefully constructed, but non-minimal derivative couplings are even more contrived \cite{nonminimalvector}. On the other hand, we will show that, within the realm of the metric-affine theories with a projective invariance considered in this work, the allowed couplings of matter fields to the curvature without Ostrogradski instabilities are much more general and, in fact, they will not be subject to any additional constraints. For all these theories, the torsion will still enter as a projective mode and, therefore, with no physical consequences. Our results thus  provide a solid and safe justification for the widely adopted condition of setting the torsion to zero, simply amounting to a choice of gauge, for a very large class of theories.

The content of the paper is organised as follows. We present in Sec. \ref{sec:I} a derivation of the field equations in a general class of metric-affine theories and apply it to the EiBI and the so-called $f(\mR)$ and $f(\mR,T)$ theories. We then generalise the formalism to include non-minimal couplings in Sec. \ref{sec:General} and we discuss the Einstein frame representation of the theories in Sec.\ref{sec:EF}. In Sec. \ref{sec:comparison} we discuss the conditions for the existence of metric compatible gauge fixing.
We conclude in Sec. \ref{sec:summary} with a summary and discussion of our results.

\section{General field equations for metric-affine theories \label{sec:I} }

For the sake of generality, in this section we derive the field equations of a general family of theories whose gravitational sector Lagrangian is a function of a metric
and the Riemann tensor of an independent connection. This will allow to set-up a general formalism and make direct contact with previous results. In order to avoid unnecessary notational complications we will assume minimally coupled fields for the moment. We will come back to this point in Sec. \ref{sec:General} to straightforwardly add non-minimal couplings complying with our requirements. Our starting point is thus the action \cite{Olmo:2012vd}
\be\label{eq:f-action}
S=\frac{1}{2\kappa^2}\int d^4x \sqrt{-g}F\left[g_{\mu\nu},{\mR^\a}_{\beta\mu\nu}(\G)\right]+S_m[g_{\mu\nu},\psi] \ ,
\ee
where $S_m$ is the matter action, $\psi$ represents collectively the matter fields, $\kappa^2$ is a constant with suitable dimensions, $g$ is the determinant of the spacetime metric $g_{\mu\nu}$,
 $F(g_{\mu\nu},{\mR^\a}_{\beta\mu\nu})$ is an arbitrary scalar function constructed with the metric and the Riemann tensor,
the field strength of the connection $\Gamma^\a_{\mu\beta}$, with components
\be
 {\mR^\a}_{ \b\m\n}=\p_\m\G_{\n\b}^\a-\p_\n\G_{\m\b}^\a+\G_{\m\l}^\a\G_{\n\b}^\l-\G_{\n\l}^\a\G_{\m\b}^\l.
\ee
We assume a symmetric metric tensor, $g_{\m\n}=g_{\n\m}$, and the usual definitions for the Ricci tensor, $\mR_{\m\n}\equiv{\mR^\r}_{\m\r\n}$, and the Ricci scalar, $\mR\equiv g^{\m\n}\mR_{\m\n}$.

The Palatini variation of the action (\ref{eq:f-action}) leads to
\be\label{eq:varS}
\d S=\frac{1}{2\k^2}\int d^4x \sqrt{-g}\left[\left(\frac{\p F}{\p g^{\m\n}} -\frac{F}{2}g_{\m\n} \right)\d g^{\m\n} + {P_\a}^{\b\m\n} \,\d{\mR^\a}_{\b\m\n}\right]+\d S_m \ .
\ee
where we have introduced the tensor
\be
{P_\a}^{\b\m\n}\equiv \frac{\p F}{\p {\mR^\a}_{\b\m\n}}\,,\label{P-tensor}
\ee
which inherits from the Riemann tensor the antisymmetry in the last two indices. Straightforward manipulations show that $\d  \mR^{\a}_{\ \b\m\n}$
can  be written as
\be\label{dRiemann}
\delta {\mR^\a}_{\beta\mu\nu}= \nabla_\mu \left(\delta \Gamma^\a_{\nu\beta}\right)-\nabla_\nu \left(\delta\Gamma^\a_{\mu\beta}\right)+2S^\lambda_{\mu\nu} (\delta\Gamma^\a_{\lambda\beta}) \ ,
\ee
where $S^\lambda_{\mu\nu}\equiv \frac{1}{2} ( \Gamma^\lambda_{\mu\nu}-\Gamma^\lambda_{\nu\mu})$ is the torsion tensor. Integrating by parts and introducing the current $J^\mu\equiv {P_\a}^{\beta\mu\nu}\d \Gamma^\a_{\n\b}$, the sector of the action that concerns the variation with respect to the connection takes the form
\ben\label{eq:step1}
I_\Gamma &\equiv&\frac{1}{\k^2}\int d^4x \sqrt{-g} {P_\a}^{\beta\mu\nu}\left(\nabla_\mu \delta \Gamma^\a_{\nu\beta} + S^\l_{\m\n}\,\delta \Gamma^\a_{\l\beta}\right)=\\
&=&\frac{1}{\k^2}\int d^4x \left\{\nabla_\mu(\sqrt{-g}J^\mu)-\left[\nabla_\mu\left(\sqrt{-g} {P_\a}^{\beta\mu\nu}\,\right) - \sqrt{-g} {P_\a}^{\beta\mu\l} S^\n_{\m\l}\right]\delta \Gamma^\a_{\nu\beta}\right\}\ ,\notag
\een
where we can isolate a boundary term as follows
\ben
I_\Gamma = \frac{1}{\k^2}\int d^4x \left\{\partial_\mu(\sqrt{-g}J^\mu)-\left[\nabla_\mu\left(\sqrt{-g} {P_\alpha}^{\beta\mu\nu}\right)-2S^\sigma_{\sigma \mu}\sqrt{-g}{P_\alpha}^{\beta\mu\nu}\right]\delta \Gamma^\alpha_{\nu\beta}\right\}\ .\label{eq:step3}
\een
Using this result, the complete variation of the action \eqref{eq:varS} reads
\begin{eqnarray}\label{eq:var2-f}
\delta S&=&\frac{1}{2\kappa^2}\int d^4x \left\{\sqrt{-g}\left(\frac{\partial F}{\partial g^{\mu\nu}} -\frac{F}{2}g_{\mu\nu} \right)\delta g^{\mu\nu}
+2\partial_\mu\left(\sqrt{-g}J^\mu\right) +\right. \\
&&+ \left.2\sqrt{-g}\left[-\tfrac{1}{\sqrt{-g}}\nabla_\mu \left(\sqrt{-g}{P_\alpha}^{\beta\mu\nu} \right)+S^\nu_{\sigma\rho}{P_\alpha}^{\beta\sigma\rho}+2S^\sigma_{\sigma\mu}{P_\alpha}^{\beta\mu\nu}\right] \delta \Gamma^\alpha_{\nu\beta}\right\}+\delta S_m \ , \nonumber
\end{eqnarray}
with associated field equations
\ben\label{eq:gmn}
\kappa^2 T_{\mu\nu}&=&\frac{\partial F}{\partial g^{\mu\nu}} -\frac{F}{2}g_{\mu\nu}  \label{eq:MetricFieldEq}\\
\kappa^2{H_\alpha}^{\nu\beta}&=&-\frac{1}{\sqrt{-g}}\nabla_\mu \left(\sqrt{-g}{P_\alpha}^{\beta\mu\nu} \right)+S^\nu_{\sigma\rho}{P_\alpha}^{\beta\sigma\rho}+2S^\sigma_{\sigma\mu}{P_\alpha}^{\beta\mu\nu} \ , \label{eq:Gamn}
\een
where  $T_{\mu\nu}\equiv-\frac{2}{\sqrt{-g}}\frac{\delta S_m}{\delta g^{\mu\nu}}$ is the energy-momentum tensor of the matter, and $ {H_\alpha}^{\nu\beta}\equiv-\frac{1}{\sqrt{-g}}\frac{\delta S_m}{\delta \Gamma^\alpha_{\nu\beta}}$ represents the coupling of matter to the connection.
As commented above, we will assume that the matter fields do not couple to the connection so that ${H_\alpha}^{\nu\beta}=0$ and discuss the effects of non-minimal couplings later on. The first of these equations represents the generalisation of Einstein's equations while the connection field equations will fix the relation between the affine and metric structures.

 In order to disentangle the symmetric part $C^\alpha_{\mu\nu}$ of the connection from the antisymmetric one, namely the torsion $S^\alpha_{\mu\nu}$, we write the connection as
 \be\label{CSdecomp}
 \Gamma^\alpha_{\mu\nu}=C^\alpha_{\mu\nu}+S^\alpha_{\mu\nu}\ .
 \ee
Then, for any vector $V_\n$ we have $ \nabla_\mu V_\nu=\partial_\mu V_\nu-C^\alpha_{\mu\nu} V_\alpha-S^\alpha_{\mu\nu} V_\alpha=\nabla_\mu^C V_\nu-S^\alpha_{\mu\nu} V_\alpha$,
and for the squared root of the metric determinant we have $\nabla_\mu \sqrt{-g} =\nabla_\mu^C \sqrt{-g}-S^\alpha_{\mu\alpha}\sqrt{-g}$, where $\nabla_\mu^C$ denotes the covariant derivative associated to the symmetric connection $C^\alpha_{\mu\nu}$.
Accordingly, equation (\ref{eq:Gamn}) reduces to
\be
\frac{1}{\sqrt{-g}}\nabla_\mu^C \left(\sqrt{-g}{P_\a}^{\b\m\n} \right)=
S^{\l}_{\m\a}{P_\l}^{\b\m\n}-S^{\b}_{\m\l}{P_\a}^{\l\m\n} \ . \label{eq:Gamn2}
\ee
After deriving the equations for a general case, we will proceed to consider the class of theories of interest for us in the present work.

\subsection{Connection Field Equations for Ricci-based Theories \label{sec:II} }

We now restrict our initial family of gravity theories to those in which only the Ricci tensor appears. A general analysis of these theories can also be found in \cite{BeltranJimenez:2017doy}.
The action that describes  these Ricci-based theories is a particular case of \eqref{eq:f-action} where the dependence on the Riemann tensor is replaced by the dependence on the Ricci tensor, namely
\be\label{eq:f-action-Ricci}
S=\frac{1}{2\kappa^2}\int d^4x \sqrt{-g}F\left[g_{\mu\nu},\mR_{\mu\nu}(\G)\right]+S_m[g_{\mu\nu},\psi] \ .
\ee
For this kind of theories, the $P-$tensor introduced in \eqref{P-tensor} takes the following form
\be
P_\a{}^{\m\b\n}=Z^{\m\r}\d_{\a\r}^{\b\n}\,,\label{PZ-tensor}
\ee
where the $Z-$tensor is defined as the derivative of the gravity Lagrangian with respect to the Ricci tensor,
\be
Z^{\m\n}\equiv \frac{\partial F}{\partial \mR_{\m\n}}  \ , \label{Z-tensor}
\ee
and $\d^{\b\n}_{\a\r}\equiv \frac12\left(\d^{\b}_{\a}\d^{\n}_{\r}-\d^{\b}_{\r}\d^{\n}_{\a}\right)$.
In terms of the $Z-$tensor, equation (\ref{eq:Gamn2}) reads
\be
\frac{1}{\sqrt{-g}}\nabla_\mu^C \left(\sqrt{-g}\,Z^{\b\r} \right)\d^{\m\n}_{\a\r}=
S^{\l}_{\m\a}Z^{\b\r}\d^{\m\n}_{\l\r}-S^{\b}_{\m\l}Z^{\l\r}\d^{\m\n}_{\a\r}\ ,\label{eq:ricci_eom}
\ee
which, once traced over $\n\a$, provides the relation
\be
-\frac{1}{\sqrt{-g}}\nabla_\mu^C \left(\sqrt{-g}\,Z^{\b\m} \right)=
\frac23 S^{\l}_{\l\m}Z^{\b\m}\d^{\m\n}_{\l\r}-S^{\b}_{\l\m}Z^{\l\m}\ .\label{eq:trace1}
\ee
Using this result, equation (\ref{eq:ricci_eom}) turns into
\be
\frac{1}{\sqrt{-g}}\nabla_\a^C \left(\sqrt{-g}\,Z^{\b\n} \right)=-\frac23\,\d^\n_\a
Z^{\b\m}S^{\l}_{\l\m} +Z^{\b\n} S^{\l}_{\l\a} - S^{\n}_{\m\a}Z^{\b\m} + S^\b_{\m\a} Z^{\m\n}\,.\label{eq:Z}
\ee
It happens that this equation can be further simplified by performing a projective transformation of the form\footnote{This shift in the connection is known as projective transformation because it leaves the geodesic paths invariant up to a redefinition of the affine parameter.}
\be\label{eq:newG}
\tilde{\G}^\l_{\m\n}=\G^\l_{\m\n}+\frac23\, A_{\m} \d^\l_\n \ ,
\ee
with the symmetric and antisymmetric part of $\tilde{\Gamma}^\lambda_{\mu\nu}$ related to those of  ${\Gamma}^\lambda_{\mu\nu}$ by
\ben\label{eq:newC}
\tilde{C}^\lambda_{\mu\nu}&=&C^\lambda_{\mu\nu} + \frac23\,\delta^\lambda_{(\nu} A_{\mu)}\,, \\
\tilde{S}^\l_{\m\n}&=&S^\l_{\m\n} + \frac23\,\delta^\lambda_{[\nu} A_{\mu]}\,. \label{eq:newS}
\end{eqnarray}
The vector $A_\mu\equiv S^\lambda_{\lambda\mu}$ is chosen in such a way that $\tilde{S}^\lambda_{\lambda\mu}=0$.
Using these new variables, equations (\ref{eq:Z}) read
\be
\frac{1}{\sqrt{-g}}\nabla_\a^{\tilde C} \left(\sqrt{-g}\,Z^{\b\n} \right)=\,{\tilde S}^{\b}_{\m\a}Z^{\m\n}-{\tilde S}^{\n}_{\m\a}Z^{\b\m}\,.\label{eq:sum}
\ee
Whenever it is possible to define the inverse of $Z^{\mu\nu}$, which we denote as $Z_{\mu\nu}$, an immediate consequence of (\ref{eq:sum}) is
\bea
\frac{1}{\sqrt{-g}}\nabla_\a^{\tilde C} \left(\sqrt{-g}\,Z^{\b\n} \right) Z_{\n\b} =0 \ .
\eea
Expanding this expression, one finds the useful relation
\bea
\tilde C^\l_{\a\l} ={\partial_\a} \big(\ln{|g|} - \frac12 \ln{|Z|}\big)\,,
\eea
where $|Z|\equiv |\det{Z_{\m\n}}|$.
This last equation can be used to simplify (\ref{eq:sum}), yielding
\be
\left(\tilde\nabla_\a  + V_\a\right)\,Z^{\b\n} = 2\,{\tilde S}^{\n}_{\a\l}Z^{\b\l}\,,\label{eq:V}
\ee
where we introduced $V_\a\equiv \frac12\partial_\a \ln\left|{\frac{Z}{g}}\right|$ and we used  $\tilde\nabla_\a$ referring to the covariant derivative of the connection $\tilde{\Gamma}^\mu_{\alpha\beta}=\tilde{C}^\mu_{\alpha\beta}+\tilde{S}^\mu_{\alpha\beta}$ defined in (\ref{eq:newG}). In order to remove the vector $V_\a$ from the above equation, we redefine the $Z-$tensor as follows
\be
Z^{\m\n} = \sqrt{\left|{\frac{g}{Z}}\right|}\,{\tilde Z}^{\m\n}\, ,\label{Weyl}
\ee
which leads to
\be
\tilde\nabla_\a {\tilde Z}^{\b\n} = 2\,{\tilde S}^{\n}_{\a\l}{\tilde Z}^{\b\l}\,.\label{eq:no-metr-up}
\ee
Guided by the standard approach of General Relativity to derive the expression of the connection, we properly sum cyclic permutations of the last equation with lowered indices,
\be
\tilde\nabla_\a \,{\tilde Z}_{\m\n} = 2\,{\tilde S}^{\l}_{\m\a}{\tilde Z}_{\l\n}\,,\label{eq:no-metr}
\ee
to finally obtain
\be
\partial_\a \,{\tilde Z}_{\m\n} + \partial_\n \,{\tilde Z}_{\a\m} - \partial_\m \,{\tilde Z}_{\n\a} =
 2\,{\tilde \G}^{\l}_{\m\a}{\tilde Z}_{[\l\n]} + 2\,{\tilde \G}^{\l}_{\a\n}{\tilde Z}_{(\l\m)} + 2\,{\tilde \G}^{\l}_{\n\m}{\tilde Z}_{[\a\l]}\,.\label{eq:master}
\ee
Symmetrising and antisymmetrising this equation in $\a\n$ one obtains, respectively,
\be
\,{\tilde C}^{\l}_{\a\n}{\tilde Z}_{(\l\m)} = \tfrac12\left(\partial_\a\,{\tilde Z}_{(\m\n)} + \partial_\n \,{\tilde Z}_{(\a\m)} - \partial_\m \,{\tilde Z}_{(\n\a)} \right) + {\tilde S}^\l_{\a\m}{\tilde Z}_{[\l\n]} + {\tilde S}^\l_{\n\m}{\tilde Z}_{[\l\a]} \ , \label{eq:KeyEqS}
\ee
\be
\tfrac12\left(\partial_\a\,{\tilde Z}_{[\m\n]} + \partial_\n \,{\tilde Z}_{[\a\m]} - \partial_\m \,{\tilde Z}_{[\a\n]} \right) = \,{\tilde C}^{\l}_{\a\m}{\tilde Z}_{[\l\n]} + {\tilde C}^\l_{\m\n}{\tilde Z}_{[\a\l]} + {\tilde S}^\l_{\a\n}{\tilde Z}_{(\m\l)} \,. \label{eq:KeyEqA}
\ee
The above two equations are key elements of this paper, as the following application shows. From the definition (\ref{Z-tensor}), it is immediate to see that if the Lagrangian only depends on the symmetric part of the Ricci tensor, then ${\tilde Z}^{\m\n}$, and its inverse ${\tilde Z}_{\m\n}$, must also be symmetric.  As a result, for such theories we have ${\tilde Z}_{[\m\n]} =0$, and Eqs. (\ref{eq:KeyEqS}) and  (\ref{eq:KeyEqA}) boil down to
\be
\,{\tilde C}^{\l}_{\a\n} = \frac12 {\tilde Z}^{\m\l}\left(\partial_\a \,{\tilde Z}_{\m\n} + \partial_\n \,{\tilde Z}_{\a\m} - \partial_\m\,{\tilde Z}_{\n\a} \right) \,,\quad\qquad {\tilde S}^\l_{\a\n}=0 \ .\quad\label{Eq:SymmCase}
\ee
This implies that the symmetric part of the connection $\tilde{\Gamma}^\alpha_{\m\n}$ coincides with the Levi-Civita connection of the (inverse) {\it auxiliary metric} ${\tilde Z}^{\m\n}$, 
while its torsion tensor ${\tilde S}^\l_{\a\n}$ vanishes. The connection $\Gamma^\lambda_{\alpha\beta}$ can thus be written as
\begin{equation}\label{eq:Gamma_sol}
\Gamma^\lambda_{\alpha\beta}=\tilde{C}^\lambda_{\alpha\beta}-\frac{2}{3} A_\alpha \delta^\lambda_\beta  \ ,
\end{equation}
with its torsion fully determined by $A_\mu$ as $S^\lambda_{\mu\nu}=\tfrac13 (\delta^\lambda_\mu A_\nu-\delta^\lambda_\nu A_\mu)$. Let us notice that the resulting connection corresponds to a particular case of the general family of connections with vector distortion introduced in \cite{Jimenez:2015fva}.

Let us now return to the definition of the Ricci tensor and its representation using the above variables
\begin{equation}\label{eq:Ricci_geral}
\mR_{\mu\nu}(\Gamma)=\mR_{\mu\nu}(\tilde{C})+\frac{2}{3}\nabla^{\tilde{C}}_{[\nu} A_{\mu]}+\nabla^{\tilde{C}}_\lambda \tilde{S}^\lambda_{\nu\mu}- \tilde{S}^\lambda_{\kappa\mu} \tilde{S}^\kappa_{\nu\lambda}\ .
\end{equation}
From this expression, it is immediate to see that with this choice of connection variables, namely $\Gamma^\alpha_{\mu\nu}$ split in a symmetric part $\tilde{C}^\alpha_{\mu\nu}$, an antisymmetric traceless part $\tilde{S}^\alpha_{\mu\nu}$, and a vector $A_\alpha$ representing the projective freedom, the symmetrised Ricci tensor takes the form
\begin{equation}\label{eq:RicciSym_geral}
\mR_{(\mu\nu)}(\Gamma)=\mR_{(\mu\nu)}(\tilde{C})- \tilde{S}^\lambda_{\kappa(\mu} \tilde{S}^\kappa_{\nu)\lambda}\ ,
\end{equation}
which is manifestly independent of the projective degrees of freedom $A_\alpha$. Theories of gravity based on the symmetric part of the Ricci tensor\footnote{In particular, any theory depending on the Ricci scalar, as $\mR=g^{\m\n}\mR_{\m\n}=g^{\m\n}\mR_{(\m\n)}$ for $g_{\m\n}$ symmetric.}, therefore, are insensitive to this projective mode, which remains undetermined by the field equations. To further reinforce this point, note that, as we have shown above, when in such theories the matter is not coupled to the connection one also finds that  ${\tilde S}^\l_{\a\n}=0$, which implies $\mR_{(\mu\nu)}(\Gamma)=\mR_{(\mu\nu)}(\tilde{C})$. The equations of motion that follow from variation of the metric, therefore, must only depend on $\tilde{C}^\alpha_{\mu\nu}$, having no trace of $A_\alpha$. 

The impossibility of determining $A_\alpha$ is simply a reflection of the fact that theories based on the symmetric part of the Ricci tensor are projectively invariant and $A_\alpha$ precisely corresponds to the projective mode that was expected to remain undetermined by the field equations. This can be clearly seen from the variation of the Ricci tensor under a general projective transformation $\Gamma^\alpha_{\mu\beta}\rightarrow \Gamma^\alpha_{\mu\beta}-\frac23A_\mu\delta^\alpha_\beta$ that is given by
\be
\mR_{\mu\nu}(\Gamma)\rightarrow\mR_{\mu\nu}(\Gamma)-\frac{4}{3}\partial_{[\mu} A_{\nu]}
\ee
where we confirm that the projective mode does not contribute to the symmetric part of the Ricci tensor. Let us notice that theories containing the full Ricci tensor will still have a pure gradient projective symmetry, i.e., they are invariant under a projective transformation where $A_\mu=\partial_\mu\theta$ with $\theta$ an arbitrary scalar. This already suggests that giving up on the projective symmetry and allowing for the general Ricci tensor will make the projective mode become a Maxwellian field. 

The arbitrary local character of the projective invariance has a direct consequence on the field equations by enforcing an off-shell constraint on the theory given by the vanishing of the trace of the connection field equations, i.e., these theories will satisfy 
\be
\frac{\delta S}{\delta \Gamma^\alpha_{\mu\beta}}\delta^\beta_{\alpha}=0.
\label{constraint1}
\ee
This constraint is unaffected by the matter sector because we have assumed it to be minimally coupled so that the corresponding hypermomentum is identically zero. We will see later how to add non-minimal couplings to the matter field such that the hypermomentum also satisfies this constraint. Before proceeding to that we will first consider two specific cases that will serve us as proxies for our discussion on the possibility of getting rid of the non-metricity by a suitable choice of the projective mode.



\subsection{EiBI theory} \label{sec:BI}

In this section, we specify the approach of the previous section to the EiBI gravity theory, which is defined by the action \cite{BIg}
\be \label{eq:SBI}
S_{\rm EiBI} = \frac{1}{\epsilon\kappa^2} \int d^4x \left[ \sqrt{-|g_{\mu\nu}+\epsilon \mR_{(\mu\nu)}(\G)| } - \lambda \sqrt{-g } \right] \ ,
\ee
where $\kappa^2 \equiv 8\pi G/c^4$ is the usual Einstein coupling constant, $\epsilon$ is a  parameter with dimensions of length squared,  the Ricci tensor is a function solely of the affine connection, which is {\it a priori} independent of the physical metric $g_{\mu\nu}$, and the parameter $\lambda$ is related to an effective cosmological constant as $\lambda=1+\epsilon \Lambda_{\rm eff}$. Note that we have made explicit the fact that the Lagrangian only depends on the symmetric part of the Ricci tensor. Recall, in this sense, that in the metric-affine formulation the Ricci tensor is not necessarily symmetric by construction, as can be seen from (\ref{eq:Ricci_geral}). The non-symmetric case will be considered elsewhere (see also the discussion at this respect in \cite{BeltranJimenez:2017doy}). 

In order to take advantage of the general approach to Ricci based theories, one just needs to identify the $Z-$tensor \eqref{Z-tensor}. Therefore, we need \eqref{eq:SBI} to match the form of expression \eqref{eq:f-action}. To this end, we identify the argument of the square root as an auxiliary metric
\begin{equation}\label{eq:qmn}
q_{\mu\nu}\equiv g_{\mu\nu} + \epsilon \mR_{(\mu\nu)}(\Gamma) \ ,
\end{equation}
which is manifestly symmetric,
and we introduce the {\it deformation matrix} 
\be
\Omega^\r{}_\n \equiv \d^{\r}{}_{\n}+\e \, g^{\r\s} \mR_{\s\n},
\ee
such that
$q_{\m\n}=g_{\m\r}\,{\O^\r}_\n $ (or $\, (q^{-1})^{\m\n}\equiv q^{\m\n}=(\O^{-1})^{\m}_{\ \s}\,g^{\s\n}$).
This allows to rewrite action \eqref{eq:SBI} as
\be\label{eq:BI_g}
S_{\rm EiBI}= \frac{1}{\k^2\epsilon}\int d^4x \sqrt{-g} \left(\sqrt{\O} -\l \right)\ ,
\ee
where $\O=\det \O^{\r}_{\ \n}$.
Thus, for the EiBI gravity we can identify $\,F(g_{\m\n}, \mR_{\m\n})=(\sqrt{\O}-\l)\,$ leading directly to the $Z-$tensor
\be
Z^{\m\n}=\sqrt{\O}\,q^{\n\m}\,.
\ee
With this result at hand, the traceless field equation for the connection \eqref{eq:Z}, assume the form
\be
\nabla_\a^C \left(\sqrt{-q} \,q^{\m\b} \right)=\sqrt{-q}\,\left[
-2 q^{\l[\m} S^{\b]}_{\a\l}
 + \left(\d^{\l}_{\a} q^{\m\b} \ -\tfrac23  \d^{\m}_{\a}q^{\l\b} \right)A_{\l} \right]\,.\label{eq:GamnBISympl}
\ee
Upon substituting the set of variables \eqref{eq:newG}, equation \eqref{eq:GamnBISympl} takes the simplified form
\be
\frac{1}{\sqrt{-q}}\nabla_\a^{\tilde C} \left(\sqrt{-q} \,\left(q^{-1}\right)^{\m\n} \right) = \tilde{S}^{\m}_{\a\l}q^{\l\n} - \tilde{S}^{\n}_{\a\l}q^{\m\l}\,.\label{eq:asymmBIb}
\ee

Following the general procedure worked out in the previous section, it is not difficult to verify that this equation can be put as
\ben
\left({\tilde\nabla}_\a + V_\a\right)\left(\sqrt{|\O|} \,q^{\m\n} \right) = 2\tilde{S}^{\n}_{\a\l}q^{\l\m} \,,\qquad V_\a\equiv -\tfrac12\partial_\a\ln{|\O|}\ .\quad
\een
This can be further simplified using the redefinition \eqref{Weyl}, which in the present case reads
\be
Z^{\m\n}=\sqrt{\O}\,q^{\n\m}= \sqrt{\O}\,\,\tilde Z^{\m\n} \quad\Rightarrow\quad \tilde Z^{\m\n}={q}^{\n\m}\, .
\ee
Note that this last result, namely that $\tilde Z^{\m\n}={q}^{\n\m}$, is consistent with our interpretation of $\tilde Z^{\m\n}$ as the inverse of the proper auxiliary metric. Finally, we are left with the analogue of \eqref{eq:no-metr}
\ben
{\tilde\nabla}_\a  q^{\m\n}  = 2\tilde{S}^{\n}_{\a\l}q^{\l\m} \ .
\een
Now, following the same steps as in the previous section, we end up with equations \eqref{Eq:SymmCase} particularised to the EiBI symmetric case
\be
\,{\tilde C}^{\l}_{\a\n} = \frac12 \,{q}^{\m\l}\left(\partial_\a \,{q}_{\m\n} + \partial_\n \,{q}_{\a\m} - \partial_\m\,{q}_{\n\a} \right) \,,\qquad {\tilde S}^\l_{\a\n}=0 \ .\quad
\ee
Actually, there is a shortcut to obtain the same result starting from \eqref{eq:asymmBIb} and noticing that we assumed the Ricci tensor to be symmetric, which implies that $q^{\m\n}$ is symmetric as well. Therefore, for consistency, the equation \eqref{eq:asymmBIb} splits into two equations
\be
\nabla_\a^{\tilde C} \left(\sqrt{-q} \,q^{\m\n} \right) = 0\,,\qquad\qquad q^{\l[\m}\tilde{S}^{\n]}_{\l\a}=0 \ .\label{eq:symmasymmBI}
\ee
Recalling that for any tensor $a_{\mu\nu}$ with determinant $a$ and inverse $a^{\mu\nu}$, the following identity 
holds
\be
\partial_\a\sqrt{|a|} = -\tfrac12 \sqrt{|a|} a_{\n\m}\partial_\a a^{\m\n}\,,
\ee
it is straightforward to prove that the first equation of \eqref{eq:symmasymmBI} is the analogue of the metricity condition for $q^{\m\n}$, leading directly to
\be
{\tilde C}^\a_{\m\n}=\frac12 q^{\a\r}\left(\p_\m q_{\r\n}+\p_\n q_{\r\m}-\p_\r q_{\m\n}\right) \,.
\ee
Concerning the second equation in \eqref{eq:symmasymmBI}, lowering free indices, one is left with $\tilde S_{[\mu\nu]\r}=0$, namely, the torsion in the new variables is symmetric in its first two indices. Therefore, it is a matter of simple algebra to conclude that $\tilde S^\a_{\m\n}=0\,$ which, in turn, implies that
\be
{S}^\l_{\m\n}=\tfrac23\, \d^\l_{[\m} A_{\n]}
\ee
as a consequence of \eqref{eq:newS}. This proves that when the field equations for the connection are (algebraically) solved, the torsion tensor still depends on the expected four projective degrees of freedom that are not fixed by the dynamics of the theory.

For completeness and for future reference, we note that the variation of (\ref{eq:BI_g}) with respect to the metric leads to
\begin{equation}
\sqrt{-q} q^{\mu\nu}-\sqrt{-g}g^{\mu\nu}=-\epsilon\kappa^2 T^{\mu\nu} \ .
\end{equation}
With algebraic manipulations, one can solve for $q_{\mu\nu}$ in terms of $g_{\mu\nu}$ and the matter as $q_{\mu\nu}=g_{\mu\alpha}{\Omega^\alpha}_\nu$, where
\begin{equation}\label{eq:Omega_BI}
\sqrt{|\Omega|}{(\Omega^{-1})^\mu}_\nu=\lambda {\delta^\mu}_\nu-\epsilon \kappa^2{T^\mu}_\nu \ .
\end{equation}
This shows that the relative deformation between the physical and auxiliary metrics is fully determined by the stress-energy tensor of the matter sources. The field equations for $q_{\mu\nu}$ can be obtained from Eq.(\ref{eq:qmn}) using the fact that $\mR_{(\mu\nu)}(\Gamma)=\mR_{(\mu\nu)}(\tilde{C})$, and take the form
\begin{equation}
{G^\mu}_{\nu}(q)=\frac{\kappa^2}{\sqrt{|\Omega|}}\left[{T^\mu}_\nu-\delta^\mu_\nu
\left(\mathcal{L}_G+\frac{T}{2}\right)\right] \ ,
\end{equation}
where $\mathcal{L}_G=(\sqrt{|\Omega|}-\lambda)/(\epsilon \kappa^2)$ represents the EiBI gravity Lagrangian, ${G^\mu}_{\nu}(q)\equiv q^{\mu\alpha}G_{\alpha\nu}(q)$, and ${T^\mu}_\nu\equiv g^{\mu\alpha}T_{\alpha\nu}$. This Einstein-like representation of the field equations for the auxiliary metric $q_{\mu\nu}$ provides an explicit example at the equations level of our general discussion in Sec.\ref{sec:EF}. This general representation was also rigorously derived for the general case in \cite{BeltranJimenez:2017doy}.

\subsection{$f(\mR)$ and $f(\mR,T)$ theories}\label{sec:f(R,T)}

Other classes of theories to which the general method of section \ref{sec:II} can be applied are the so-called $f(\mR)$ theories \cite{Nojiri:2010wj,DeFelice:2010aj,Olmo:2011uz}, with $\mR\equiv g^{\mu\nu}\mR_{\mu\nu}(\Gamma)$. For the same price we can also consider the $f(\mR,T)$ theories, with $T$ representing the trace of the stress-energy tensor of the matter fields, that have recently received some attention in the literature \cite{Harko:2011kv,Momeni:2011am,Houndjo:2011tu,Alvarenga:2013syu,Yousaf:2016lls}. Furthermore, this will be our first step towards including non-minimal couplings, case that will be fully addressed in the next section. In both cases, the dependence on $\mR$ guarantees the projective invariance. For these theories, the action is the same as \eqref{eq:f-action} provided the function $F$ is replaced by the corresponding $f$ (from now on we use just $f$ to denote both $f(\mR)$ and $f(\mR,T)$ theories). The restricted dependence of the function $F$ specifies the form of the $Z-$tensor \eqref{Z-tensor} as follows
\be\label{Pf}
Z^{\mu\nu}=  f_\mR\,  g^{\m\n} \ ,
\ee
where $f_\mR\equiv\partial_\mR f$ (for both, $f(\mR)$ and $f(\mR,T)$ theories). Plugging this expression into \eqref{eq:ricci_eom}, the field equations for the connection become
\be
\tfrac{1}{\sqrt{-g}}\nabla^C_\m \left(\sqrt{-g}f_\mR \,g^{\s\b} \right)\d^{\m\n}_{\a\s} =
\tfrac12  f_\mR  \left(S^{\l}_{\l\a} g^{\n\b} - S^{\b}_{\a\l} g^{\n\l} - S^{\n}_{\s\a} g^{\s\b}\,  \right)\,.
\ee
which can be further simplified using its trace equation and the shift transformation of the connection \eqref{eq:newG} to end up with the analogue of \eqref{eq:sum}, namely
\be
\tfrac{1}{\sqrt{-g}}\nabla^{\tilde{C}}_\a \left(\sqrt{-g}f_\mR\,g^{\m\n} \right) = 2  f_\mR \tilde{S}^{[\m}_{\l\a}\, g^{\n]\l} \ .
\ee
This equation makes it manifest that, as a consequence of the symmetry of the metric, both sides must identically vanish, yielding
\ben\label{eq:symm2}
\tfrac{1}{\sqrt{-g}}\nabla_\alpha^{\tilde{C}}(\sqrt{-g}\,f_\mR\, g^{\beta\nu}) &=& 0\\
\tilde S_{\a\n\b} -\tilde S_{\b\n\a} = 0 \q&\Leftrightarrow&\q \tilde S^{\l}_{\m\n} = 0\ . \label{eq:asymm2}
\een
The general procedure of section \ref{sec:II}, allows to identify the auxiliary metric in the form
\be
h^{\m\n}=\tilde Z^{\m\n} = f_\mR^{-2} Z^{\m\n} = \frac{1}{f_\mR} g^{\m\n}\,,\label{Aux-Metric-f}
\ee
leading to the conclusion \eqref{Eq:SymmCase} that the symmetric part of the shifted connection coincides with the Christoffel symbol associated to the auxiliary metric, that is,
\be
\tilde C^{\a}_{\m\n}=\tfrac{1}{2}\, h^{\a\rho}\left(\p_\m h_{\rho\n}+\p_\n h_{\rho\m}-\p_\rho h_{\m\n}\right)\,.
\ee
We thus see that the connection is generated by a metric that is conformally related to the spacetime metric so that the deformation matrix is proportional to the identity matrix $\Omega^\alpha{}_\beta=f_\mR \delta^\alpha{}_\beta$. As we will show later, this simple relation between the two metrics has an important impact on the existence of a metric-compatible projective gauge.

\section{General projectively invariant non-minimal couplings}\label{sec:General}

In the previous section we have considered a type of theories where matter fields can couple non-minimally through the trace of the energy-momentum tensor and, nevertheless, the same results found for minimally coupled fields apply to this case as well. The specific coupling through the trace of the energy-momentum tensor does not have any special structure and we will show in the following that it is nothing but a particular case of a larger class of theories with non-minimally coupled matter fields.

Let us then consider general affine theories in the presence of matter fields with non-minimal couplings. We will restrict the couplings as to maintain the projective invariance in a very simple way, namely by allowing couplings to the curvature only through the symmetric part of the Ricci tensor so that we will consider theories generally described by the following action
\begin{equation}\label{eq:F-nonminimal}
 S=\frac{1}{2\kappa^2}\int d^4xF(g_{\mu\nu}, \mR_{(\mu\nu)}(\Gamma),\psi,\partial\psi)
\end{equation}
with $F$ an arbitrary function of the metric, the symmetric part of the Ricci tensor and the matter fields collectively denoted by $\psi$. Note that we only allow up to first derivatives of the matter fields and we have intentionally spelled out that such derivatives are indeed partial derivatives and not covariant ones. For bosonic fields this is actually the natural situation. If $\psi$ corresponds to a scalar field $\varphi$, the covariant derivatives reduce to partial derivatives (assuming we are dealing with true scalar fields and not scalar densities) $\nabla_\mu\phi=\partial_\mu\phi$. Thus, we can include non-minimimal couplings like e.g. $\mR^{\mu\nu}\partial_\mu\varphi\partial_\nu\varphi$ that would lead to Ostrogradski instabilities in the metric formalism. For vector fields $A_\mu$ we assume that all derivatives enter through the corresponding field strength $F_{\mu\nu}=\partial_\mu A_\nu-\partial_\nu A_\mu$. We adopt the standard definition of the field strength as the exterior derivative of the vector field, which naturally extends our considerations to higher $p$-forms. Sometimes the field strength in curved spacetime is defined by promoting partial derivatives to covariant ones in the flat spacetime definition of $F_{\mu\nu}$ in which case an explicit coupling to the torsion arises (that would not be allowed by our requirements). In this case we can include interactions of the form $\mR^{\mu\nu}A_\mu A_\nu$ or $\mR^{\mu\nu}F_{\mu\alpha}F_\nu{}^\alpha$ that, again, would give rise to ghost-like instabilities in the purely metric framework. Things are more subtle when including fermions into the picture since they couple directly to the spin connection with minimal interactions. Fortunately, the minimally coupled Dirac Lagrangian still exhibits a projective symmetry and, thus, our analysis can be consistently extended to the case of fermionic fields. However, adding non-minimal couplings for fermions with our requirements must be done with more care than with bosonic fields. For simplicity in the derivation of our main result we will assume that the independent connection $\Gamma^\alpha_{\mu\nu}$ only appears through the Ricci tensor. In that case, the variation of the action with respect to the connection leads to 
\begin{equation}
\delta S=\frac{1}{2\kappa^2}\int d^4x \sqrt{-g}\frac{\partial F}{\partial \mR_{(\mu\nu)}}\delta \mR_{(\mu\nu)}(\Gamma) \ ,
\end{equation}
and the manipulations of Sec. \ref{sec:II} recover the same results as we found there for minimally coupled theories. In particular, we see that we can introduce the auxiliary metric $\Zt_{\mu\nu}$ defined by
\be
\sqrt{-g}\frac{\partial F}{\partial \mR_{(\mu\nu)}}\equiv\sqrt{-\Zt}\Zt^{\mu\nu}
\ee
so that the resolution of the connection field equations follows identically and, thus, the connection will be given by the Levi-Civita connection of the auxiliary metric $\Zt_{\mu\nu}$ up to the projective mode. Obviously, similarly to the minimally coupled case, the auxiliary metric itself depends on all the arguments of the function $F$ among which we find the Ricci tensor. The crucial point is once again that the metric field equations
\be
\frac{\partial F}{\partial g_{\mu\nu}}-\frac12F g_{\mu\nu}=\kappa^2T_{\mu\nu}
\ee
allow to algebraically obtain the symmetric part of the Ricci tensor in terms of the spacetime metric and the matter fields. This is where the differences between minimally and non-minimally coupled theories arise. If the couplings are restricted to be minimal, then the matter fields only appear on the RHS of the above equation and the symmetric part of the Ricci tensor will be given in terms of $g_{\mu\nu}$ and the usual energy-momentum tensor of the corresponding matter fields. For non-minimally coupled fields, the LHS will also depend on the matter fields so that, although we will still be able to express the Ricci tensor in terms of $g_{\mu\nu}$ and the matter fields, the latter will appear in arbitrary combinations, not necessarily through the energy-momentum tensor. 

Finally, it is important to note that the hypermomentum $H_\alpha^{\mu\beta}$ of the matter fields will not vanish and, therefore, it will contribute non-trivially to the connection field equations. As we have discussed, this fact can be easily handled and the resulting effect will be a different dependence of the auxiliary metric on the matter fields. Moreover, the existence of the projective symmetry in the non-minimal couplings will guarantee that the off-shell constraint given in (\ref{constraint1}) will not be violated, i.e., we will have $H_\alpha^{\mu\alpha}=0$. It is sometimes argued that the projective symmetry in metric-affine theories (mainly within the context of the Einstein-Hilbert action as in \cite{Hehl1978}) leads to inconsistencies in the field equations when introducing matter fields. However, as discussed for instance in \cite{BeltranJimenez:2017doy}, even in the presence of non-minimal couplings, inconsistencies do not necessarily arise and our results here provide an explicit realisation of this general argument by simply imposing the same projective symmetry in the matter sector, which is in fact a natural manner of introducing non-minimal couplings within the framework of these theories.

\section{Einstein frame representation}\label{sec:EF}

In the previous sections we have worked with the field equations directly to show that the connection is given by the Levi-Civita connection of the auxiliary metric and the torsion only enters as a projective mode which can therefore be gauged away. We will now re-obtain the same results from a different approach by showing the existence of an {\it Einstein frame} representation for generic gravity Lagrangians which depend on the symmetric part of the Ricci tensor, including the non-minimal couplings discussed in the precedent section. This naturally extends the Einstein frame representation constructed in \cite{BeltranJimenez:2017doy} for the case of minimally coupled fields to the case of non-minimally coupled fields and the full discussion given in that reference will be valid for the more general theories under consideration here. The Einstein frame is achieved by first performing a Legendre transformation so that the action is written in the equivalent form 
\begin{equation}\label{eq:EF0}
S=\frac{1}{2\kappa^2}\int d^4x \sqrt{-g}\left[F\left(g_{\mu\nu},\Sigma_{\mu\nu},\psi,\partial\psi\right)+\frac{\partial F}{\partial \Sigma_{\mu\nu}}\big(\mR_{(\mu\nu)}-\Sigma_{\mu\nu} \big)\right]+S_m[g_{\mu\nu},\psi,\partial\psi] \ ,
\end{equation}
where $\Sigma_{\mu\nu}$ is an auxiliary field. 
One can easily verify that variation with respect to $\Sigma_{\mu\nu}$ leads to $\mR_{(\mu\nu)}(\Gamma)=\Sigma_{\mu\nu}$ (provided $F$ is a non-linear function of $\Sigma_{\mu\nu}$), which inserted back into the action recovers the theory (\ref{eq:f-action-Ricci}) in the Ricci symmetric case. A key reason to introduce the above action is that, unlike (\ref{eq:F-nonminimal}), this is linear in the Ricci tensor $\mR_{(\mu\nu)}(\Gamma)$. We can now introduce the field redefinition given by
\begin{equation}\label{eq:ZmnZtmn}
\sqrt{-\Zt}\Zt^{\mu\nu}=\sqrt{-g}\frac{\partial F}{\partial \Sigma_{\mu\nu}} \ ,
\end{equation}
that allows to express $\Sigma_{\mu\nu}$ in terms of the new field $\Zt_{\mu\nu}$, the spacetime metric $g_{\mu\nu}$ and the matter field, i.e., $\Sigma_{\mu\nu}=\Sigma_{\mu\nu}(\Zt_{\mu\nu},g_{\alpha\beta},\psi,\partial\psi)$. In terms of the new field $\tilde{Z}_{\mu\nu}$, the action can be written as
\begin{equation}\label{eq:EF1}
S=\frac{1}{2\kappa^2}\int d^4x \left[ \sqrt{-\tilde{Z}}\tilde{Z}^{\mu\nu}\mR_{(\mu\nu)}(\Gamma)-\sqrt{-g}V(g_{\alpha\beta},\tilde{Z}_{\alpha\beta},\psi,\partial\psi)\right]+S_m[g_{\mu\nu},\psi,\partial\psi] \ ,
\end{equation}
where the {\it potential} $V(g_{\alpha\beta},\tilde{Z}_{\alpha\beta},\psi,\partial\psi)$ is defined as 
\begin{equation}
V(g_{\alpha\beta},\tilde{Z}_{\alpha\beta},\psi,\partial\psi)=\frac{\partial F}{\partial \Sigma_{\mu\nu}}\Sigma_{\mu\nu}-F \ .
\end{equation}
Here, the field $\Sigma_{\mu\nu}$ must be replaced by its expression obtained from solving (\ref{eq:ZmnZtmn}). This is the step where the differences between minimally and non-minimally coupled theories arise since in the minimally coupled case this potential does not depend on the matter fields. On the other hand, we can see from the equivalent action (\ref{eq:EF1}) that the field $\tilde{Z}_{\mu\nu}$ has now the standard Einstein-Hilbert kinetic term and the spacetime metric $g_{\mu\nu}$ only enters algebraically, i.e. it is an auxiliary field that can be integrated out by solving its field equations. Since now the potential $V$ also depends on the matter fields, the solution for $g_{\mu\nu}$ does not need to be a function of the energy-momentum tensor as it occurs in the minimally coupled case. The corresponding solution will thus provide $g_{\mu\nu}=g_{\mu\nu}(\Zt_{\alpha\beta},\psi,\partial\psi)$ that can be plugged in the action to finally obtain
\begin{equation}\label{eq:EF2}
S=\frac{1}{2\kappa^2}\int d^4x\sqrt{-\tilde{Z}}\tilde{Z}^{\mu\nu}\mR_{(\mu\nu)}(\Gamma)+\tilde{S}_m[\Zt_{\mu\nu},\psi,\partial\psi] \ ,
\end{equation}
where $\tilde{S}_m$ stands for the modified matter sector resulting from integrating the spacetime metric out. Since the auxiliary metric $\Zt_{\mu\nu}$ now has the usual Einstein-Hilbert kinetic term in the Palatini formalism, we can now use the extensive machinery developed for this simple action and we can immediately conclude that the connection will be nothing but the Levi-Civita connection of $\Zt_{\mu\nu}$ up to the projective mode \cite{Hehl1978}, which remains as a pure gauge freedom (see e.g \cite{Julia:1998ys,Dadhich:2010xa}). The only difference now is that we have generated new matter interactions. 

We thus recover the results obtained above when working at the field equations level. In this representation we can see again the difference between minimally and non-minimally coupled cases, namely: while in the former the matter fields only enter through the usual energy-momentum tensor in the relation of the two metrics, the latter gives rise to a more general dependence on the matter fields, not necessarily through the energy-momentum tensor. A very straightforward application of our result concerns the extensions of the 
$f(\mR,T)$ theories proposed in \cite{Odintsov:2013iba} to include general couplings of the Ricci tensor and the energy-momentum tensor so that the action depends on an arbitrary function $f(\mR,T,\mR^{\mu\nu}T_{\mu\nu})$. Although these theories give rise to interesting cosmological phenomenology in their metric formulation, it was shown in \cite{Ayuso:2014jda} that the additional coupling to the Ricci tensor inevitably introduces ghost-like instabilities when considered as universal. We see however that these theories can be rendered ghost-free in their metric-affine formulation according to the results obtained here, since they correspond to a particular case of the general matter couplings considered above.

\section{Projective gauge fixing and existence of a metric gauge}\label{sec:comparison}

In the previous sections we have extensively studied the consistency of theories with a projective symmetry, including non-minimal couplings. Due to this symmetry we have encountered the expected presence of four gauge degrees of freedom in the solution of the connection encoded in the vector $A_\m=S^\lambda_{\lambda\mu}$. In this section we will exploit this symmetry and discuss the existence of special gauge choices. In particular, we will analyse the conditions under which a gauge without non-metricity is possible.

Let us first consider the case of $f(\mR)$ and $f(\mR,T)$ theories already explored above. In those theories, the projective modes can be used to obtain different representations of the same theory in such a way that one can have $i)$  a torsionless theory with nontrivial non-metricity or $ii)$ a metric compatible theory with torsion. Obviously, both representations being simply two different gauge choices are physically equivalent, but the geometrical frameworks differ\footnote{Let us comment that equivalent representations in terms of different geometrical entities of the same gravity theory already exist in the literature. Perhaps, the paradigmatic example is the equivalence between General Relativity, with gravity represented by the curvature of the Levi-Civita connection, and its teleparallel equivalent, where there is no curvature and gravity is ascribed to the torsion of the Weitzenb\"ock connection.}. In order to explicitly show this, we will consider the general solution for the connection (\ref{eq:Gamma_sol}), which consists of the piece compatible with the auxiliary metric, that in the present case is given by $h_{\mu\nu}=f_\mR \,g_{\mu\nu}$, plus the projective mode. By taking the $\Gamma$-covariant derivative of this expression we then have $\nabla^\Gamma_\alpha h_{\mu\nu}=\frac43 A_\alpha h_{\mu\nu}$ so that the covariant derivative of the spacetime metric is given by
\begin{equation}\label{eq:QamnfR}
\nabla^\Gamma_\alpha g_{\mu\nu}=\left(\tfrac{4}{3}A_\alpha-\partial_\alpha \log f_\mR \right)g_{\mu\nu} \ .
\end{equation}
In the study of $f(\mR)$ and $f(\mR,T)$ theories \`{a} la Palatini, it is common to set the torsion to zero ($A_\alpha=0$) at the onset of the analysis, which implies that  the non-metricity tensor
$Q^\Gamma_{\alpha\mu\nu}\equiv\nabla^\Gamma_\alpha g_{\mu\nu}$ takes the value $Q^\Gamma_{\alpha\mu\nu}=-g_{\mu\nu}\,\partial_\alpha \log f_\mR $. However, the above equation (\ref{eq:QamnfR}) shows that the choice $A_\alpha=\frac{3}{4}\partial_\alpha \log f_\mR$ leads to $Q^\Gamma_{\alpha\mu\nu}=0$, which implies a metric-compatible connection, but with torsion. An explicit analysis of the resulting field equations puts forward that these two representations are physically equivalent, as a consequence of the freedom in the choice of $A_\alpha$ --see \cite{Olmo:2011uz} for details.

Let us now investigate whether the Born-Infeld gravity theory exhibits similar features. The most commonly (if not uniquely) studied case is the torsionless gauge (frequently without explicit mention to that being a gauge choice of the projective mode). We want to study here it this theory could also admit a metric-compatible representation with some nontrivial torsion. We are now ready to check this by explicitly computing $\nabla^\Gamma_\alpha\,g_{\mu\nu}$ having in mind that, unlike in the $f(\mR)/f(\mR,T)$ case, the particular relation between $g_{\mu\nu}$ and $q_{\mu\nu}$ is no longer conformal but of the form $q_{\mu\nu}=g_{\mu\alpha}\,{\Omega^\alpha}_\nu$, with ${\Omega^\alpha}_\nu$ satisfying (\ref{eq:Omega_BI}). In this case, we find
 \begin{equation}\label{eq:QamnfR}
\nabla^\Gamma_\alpha \,g_{\mu\nu}=\frac{4}{3}A_\alpha g_{\mu\nu}+\nabla^{\tilde{C}}_\alpha g_{\mu\nu} \ .
\end{equation}
The computation of $\nabla^{\tilde{C}}_\alpha g_{\mu\nu}$ can be efficiently done by employing the relation $g_{\mu\nu}=q_{\mu\beta}{(\Omega^{-1})^\beta}_\nu ={(\Omega^{-1})_\nu}^\beta q_{\beta\mu}\,$, and the fact that $\nabla^{\tilde{C}}_\alpha q_{\mu\nu}=0$, leading to 
 \begin{equation}\label{eq:QamnBI0}
\nabla^\Gamma_\alpha \,g_{\mu\nu}=\frac{4}{3}A_\alpha g_{\mu\nu}+2{\Omega^\lambda}_\beta\left[g_{\lambda(\mu}\nabla^{\tilde{C}}_\alpha {(\Omega^{-1})^\beta}_{\nu)}\right]=\frac{4}{3}A_\alpha g_{\mu\nu}-2\left[\nabla^{\tilde{C}}_\alpha{\Omega^\lambda}_\beta\right] g_{\lambda(\mu}{(\Omega^{-1})^\beta}_{\nu)} \,
\end{equation}
that can be more compactly written as\footnote{We drop the explicit symmetrisation because one can show that the object on the RHS is symmetric by construction \cite{BeltranJimenez:2017doy}.}
 \begin{equation}\label{eq:QamnBI}
\nabla^\Gamma_\alpha \,g_{\mu\nu}=\left[\frac43A_\alpha\delta^\lambda{}_\nu
-\nabla^\Gamma_\alpha\big(\log\Omegah\big)^\lambda{}_\nu\right]g_{\mu\lambda}
\end{equation}
which nicely generalises the expression (\ref{eq:QamnfR}) to the case of Born-Infeld theories. Furthermore, since we only used that the two metrics are related by means of the deformation matrix $\Omega^\mu{}_\nu$ to obtain the above expression, it is in fact completely general for the class of theories considered in the precedent sections built in terms of the symmetric part of the Ricci tensor. Let us also notice that $\nabla^\Gamma_\alpha\Omega^\mu{}_\nu=\nabla^{\tilde{C}}_\alpha\Omega^\mu{}_\nu$  because $\Omega^\mu{}_\nu$ is a $(1,1)$-tensor and, then, one can easily show that the $A_\alpha-$piece of the $\Gamma-$covariant derivative identically vanishes. As it is apparent from (\ref{eq:QamnBI}), those theories for which the deformation matrix is proportional to the identity, i.e., the two metrics are conformally related, admit a gauge such that the non-metricity tensor vanishes. In the general case where the deformation matrix has a more involved structure (like in Born-Infeld theories for instance) one cannot, in general, find a gauge with vanishing non-metricity. 

Let us obtain more explicit conditions on the theories that permit a metric-compatible gauge. For that, let us start from the condition
\be
\frac43A_\alpha\delta^\lambda{}_\nu
-\nabla^\Gamma_\alpha\big(\log\Omegah\big)^\lambda{}_\nu=0.
\label{eq:condition1}
\ee
The existence of solutions for this equation will guarantee the existence of a gauge without non-metricity. If we take the trace with respect to $\lambda$ and $\nu$ we obtain
\be
A_\alpha=\frac{3}{16}\partial_\alpha\log \Omega
\ee
where we remind that $\Omega=\det\Omegah$ and we have used that $\text{ Tr}\log\Omegah=\log\det\Omegah$. We can now insert this expression in (\ref{eq:condition1}) to rewrite it purely in terms of the deformation matrix as
\be
\nabla^\Gamma_\alpha\Big[\log\Omega^\lambda{}_\nu-\frac14(\log\Omega) \delta^\lambda{}_\nu\Big]=0.
\label{eq:cond2}
\ee
Since the deformation matrix is fully determined by the form of the action, this equation gives the desired condition for the existence of a metric gauge. We will not analyse here in detail the most general theories satisfying these equations. In fact, it is conceivable that some theories exhibit solutions of the field equations compatible with (\ref{eq:cond2}) so that the non-metricity can only be removed on some specific solutions, but not in general, as we will illustrate below for a particular case. However, the condition (\ref{eq:cond2}) immediately shows that theories giving rise to a deformation matrix proportional to the identity, like those where the curvature only enters through the Ricci scalar $\mR$, admit a gauge without non-metricity. For those theories one can entirely trade the torsion and the non-metricity with the corresponding equivalent representations of the theory.

In order to gain some insight on the conditions needed to obtain a metric compatible gauge for the general case, we will now consider a scenario in which all off-diagonal terms in ${\Omega^\lambda}_\beta$ vanish, leaving only a diagonal object of the general form ${\Omega^\lambda}_\beta=\omega_{(\beta)}\, \delta^\lambda_\beta$, with up to four different functions $\omega_{(\beta)}$ (no summation over repeated indices). This is also a physically relevant case as many applications involving perfect fluids, scalar fields, electric fields, etc. in  spherically symmetric or cosmological scenarios typically involve a diagonal deformation matrix ${\Omega^\lambda}_\beta$ \cite{BeltranJimenez:2017doy}.
For instance, considering a perfect fluid of density $\rho$ and pressure $P$ as matter source in homogeneous and isotropic cosmological models, one finds ${T^\mu}_\nu=\text{diag}[-\rho,P,P,P]$, whose symmetries allow for a deformation matrix of the form ${\Omega^\mu}_\nu=\text{diag}[\omega_0,\omega_1,\omega_1,\omega_1]$, with $\omega_0$ and $\omega_1$ being functions of the cosmic time $t$. In static stellar models we find a similar decomposition but with $\omega_0$ and $\omega_1$ being functions of the radial coordinate. In black hole scenarios with electric charge $q$, we find ${T^\mu}_\nu=\frac{q^2}{8\pi r^4}\text{diag}[-1,-1,+1,+1]$, whose symmetries can again be assumed for the deformation matrix, namely  ${\Omega^\mu}_\nu=\text{diag}[\omega_0,\omega_0,\omega_1,\omega_1]$, with $\omega_0$ and $\omega_1$ now being functions of the radial coordinate \cite{Olmo:2016tra}. A similar structure appears in the case of certain anisotropic fluids \cite{Bejarano:2017fgz,Olmo:2015axa}. For a static, spherically symmetric scalar field, one has ${T^\mu}_\nu=\text{diag}[T_0,T_1,T_0,T_0]$ and so one can assume ${\Omega^\mu}_\nu=\text{diag}[\omega_0,\omega_1,\omega_0,\omega_0]$, with $\omega_0$ and $\omega_1$ being again functions of the radial coordinate \cite{Afonso:2017aci}. Note that these algebraic properties are quite general and are also valid in theories different from the EiBI gravity model. 


In the diagonal cases, we then have that (\ref{eq:QamnBI}) reduces to
 \begin{equation}\label{eq:QamnBI3}
\nabla^\Gamma_\alpha \,g_{\mu\nu}=\left(\tfrac{4}{3}A_\alpha - \partial_\alpha \ln \omega_{(\nu)}\right) g_{\mu\nu} \ . 
\end{equation}
According to this expression, for the simplest set ups, namely, those with diagonal ${T^\mu}_\nu$ and ${\Omega^\mu}_\nu$, it is not possible in general to set the right-hand side of (\ref{eq:QamnBI3}) to zero (vanishing non-metricity) by specifying the form of $A_\alpha$ due to the freedom in the choice of the four gradients $\partial_\alpha\omega_{(\nu)}$. Only when ${\Omega^\mu}_\nu$ represents a conformal transformation (which is the case of theories where the curvature only enters through the Ricci scalar like in $f(\mR)$ and $f(\mR,T)$ theories) is it possible to completely remove the non-metricity in favour of a specific type of torsion. In all the examples mentioned above, where  the functions $\omega_{(\nu)}$ only depend on one coordinate (time or radius), the vector $A_\alpha$ can only absorb one of the gradients of the $\omega_{(\nu)}$ and, therefore, the resulting theories possess genuine non-metricity.  

Summarising, the results for this simple scenario put forward that only in the case in which the four $\omega_{(\nu)}$ are the same function one can use the projective gauge freedom to get rid of all the non-metricity terms. This case in fact corresponds to a conformal transformation between the physical and the auxiliary metrics, which is characteristic of theories where the deformation matrix is proportional to the identity, as it happens for theories only depending on the Ricci scalar $\mR$. In the EiBI theory, however, this does not happen in general, except for a restricted class of solutions supported by very specific matter sources like e.g. the case of a cosmological constant. This illustrates our discussion for the general case of how the non-metricity can be removed for some solutions even if the general deformation matrix of the theory does not identically fulfils the condition (\ref{eq:cond2}).

\section{Summary and conclusion}\label{sec:summary}

In this work we have developed a general formalism for metric-affine theories of gravity with a projective invariance that is implemented by considering Lagrangians where the connection only enters through the symmetric part of the Ricci tensor. We have naturally extended the results of \cite{BeltranJimenez:2017doy} to the case of non-minimally coupled fields. In order to comply with the projective symmetry, the non-minimal couplings are again imposed to contain only the symmetric part of the Ricci tensor. For this very large class of theories we have obtained the general solution for the connection. We have shown that the torsion only appears as a projective mode and, thus, can be gauged away. On the other hand, the symmetric part of the connection (carrying the physical information) is given by the Levi-Civita connection of an auxiliary metric that is non-trivially related to the spacetime metric and the matter fields. An important property that we have clarified is that the presence of non-minimally coupled matter fields make the dependence of the auxiliary metric on them more general than in the minimally coupled case, where matter fields only enter through the energy-momentum tensor. This can in fact have interesting phenomenological consequences, which will be explored elsewhere.

Finally, we have discussed the existence of a projective gauge choice without non-metricity.
We have shown that such a gauge always exists for theories where the two metrics are conformally related, essentially reducing this family of theories to those where only the Ricci scalar appears. On the other hand, if the two metrics are not conformally related it is, in general, not possible to impose a gauge without non-metricity. We have explicitly shown our results for the case of $f(\mR)$ and $f(\mR,T)$ theories as examples of theories admitting a metric-compatible representation and Eddington-inspired-Born-Infeld as an example where such a representation is not possible so that non-metricity is, therefore, an  intrinsic, genuine property of such theories. 

We would like to remark an interesting physical difference between the two families of theories obtained according to the existence or not of a gauge with vanishing non-metricity.
From the two frames existing in these theories, one can easily see that gravitons propagate on the auxiliary metric\footnote{The relevance of the auxiliary metric for the propagation of the tensor modes was previously shown in \cite{Jimenez:2015caa} for general Palatini theories and in \cite{Bazeia:2015zpa} for braneworld scenarios.}, while photons propagate on the spacetime metric (see the discussion on this point in \cite{BeltranJimenez:2017doy}) so that, even though they both are massless particles, their trajectories can differ.
As we have discussed, the metric gauge exists for theories where the two metrics are conformally related and, since null geodesics are conformally invariant, these theories are characterised by the universality of the propagation of massless particles. On the other hand, theories not allowing the gauge with vanishing non-metricity will typically be characterised by a different propagation of photons and gravitons even if they are both massless particles.

Though this work has focused on the properties of projectively invariant theories, the master equations (\ref{eq:KeyEqS}) and  (\ref{eq:KeyEqA}) also contemplate cases without this symmetry. The physical implications of the breakdown of this symmetry and the role of the pure tensorial part of the torsion, $\tilde{S}^\lambda_{\alpha\beta}$, in cosmological and astrophysical scenarios would be interesting to pursue. On the other hand, the noticed relation between the general solution of the connection including the projective mode and the family of geometries with vector distortion introduced in \cite{Jimenez:2015fva} also opens an interesting possibility of partially unleashing the torsion or the connection with a general vector mode, this time dynamical. This would be at the expense of breaking the projective invariance, but with potentially interesting cosmological scenarios as those studied in \cite{Jimenez:2016opp}. Irrespective of the specific route, our results encourage to further explore metric-affine theories as those considered here and possible extensions.

\section*{Acknowledgments}

V.I.A. is supported by the postdoctoral fellowship CNPq-Brasil/PDE No.~234432/2014-4 and by Federal University of Campina Grande (Brazil). C. B. is funded by the National Scientific and Technical Research Council (CONICET). JBJ acknowledges the financial support
of A*MIDEX project (n ANR-11-IDEX-0001-02) funded by the Investissements d'Avenir French Government program, managed by the French National Research Agency (ANR), MINECO (Spain) projects FIS2014-52837-P, FIS2016-78859-P (AEI/FEDER) and Consolider-Ingenio MULTIDARK CSD2009-00064. G.J.O. is supported by a Ramon y Cajal contract  and the Spanish grant FIS2014-57387-C3-1-P (MINECO/FEDER, EU). This work has also been supported by the i-COOPB20105 grant of the Spanish Research Council (CSIC), the Consolider Program CPANPHY-1205388, the Severo Ochoa grant SEV-2014-0398 (Spain), and the CNPq (Brazilian agency) project No.301137/2014-5. V.I.A. and C.B. thank the Department of Physics of the University of Valencia for their hospitality during the elaboration of this work. This article is based upon work from COST Action CA15117, supported by COST (European Cooperation in Science and Technology).

\end{document}